%% file: paper.tex
\documentclass[useAMS,usenatbib]{mn2e}

\usepackage[dvips]{graphicx}
\usepackage{epsfig,amssymb,amsmath,color}
\newcommand{\beq}{\begin{equation}}
\newcommand{\eeq}{\end{equation}}
\newcommand{\beqn}{\begin{eqnarray}}
\newcommand{\eeqn}{\end{eqnarray}}

\long\def\symbolfootnote[#1]#2{\begingroup%
\def\thefootnote{\fnsymbol{footnote}}\footnote[#1]{#2}\endgroup}

\title[On the interpolation of calibration solutions obtained in radio interferometry]{On the interpolation of calibration solutions obtained in radio interferometry}
\author[Sarod Yatawatta]{Sarod Yatawatta$^{1}$\thanks{E-mail:
yatawatta@astron.nl}\\
$^{1}$ASTRON, The Netherlands}
\begin{document}

\date{\today}

\pagerange{\pageref{firstpage}--\pageref{lastpage}} \pubyear{2012}

\maketitle

\label{firstpage}

\begin{abstract}
Full polarimetric radio interferometric calibration is performed by estimating 2 by 2 Jones matrices representing instrumental and propagation effects. The solutions obtained in this way differ from the true solutions by a 2 by 2 unitary matrix ambiguity. This ambiguity is common to all stations for which a solution is obtained but it is different for solutions obtained at different time and frequency intervals. Therefore, straightforward interpolation of solutions obtained at different time and frequency intervals is not possible. In this paper, we propose to use the theory of quotient manifolds for obtaining correct interpolants that are immune to unitary matrix ambiguities.
\end{abstract}


\section{Introduction}
Most modern radio interferometers have dual polarized feeds and therefore, the use of the matrix measurement equation \citep{HBS} gives a compact and accurate description of their operation. Calibration of such an interferometer essentially boils down to the estimation of Jones matrices of size $2 \times 2$ with complex entries. As shown by \cite{H4}, the solutions acquired for the Jones matrices will only be equivalent to the true solution upto a unitary matrix ambiguity. 

This ambiguity would not hinder further processing of data because it cancels out during correction of the data using the obtained solutions. However, the ambiguities do prevent us from using the solutions for further modeling of instrumental effects (e.g., beam shape  \citep{SBYBeam}) and effects due to the propagation medium such as the ionosphere \citep{Int09}. Moreover, interferometers operating at low frequencies have a wide field of view and calibration has to be performed along hundreds of directions in an efficient manner \citep{Kaz2}. Along each direction, we would have an ambiguity which is independent of other directions.

In this paper, we present a method of interpolation of the calibration solutions (Jones matrices) and we consider the case where each solution is affected by any unknown unitary ambiguity. We consider the simplest case of interpolation to present our method: Finding the mean of a given set of solutions which can also be extended to interpolation with weighted averaging. The uses of interpolation or averaging are numerous. First, averaging of solutions obtained at adjacent time and frequency intervals provides us with a robust estimate of calibration solutions especially under noisy situations. Moreover, interpolation reduces the number of data points that needs to be visualized. This is important when we have solutions over hundreds of directions in the sky at a number of time and frequency intervals. Interpolation also provides us solutions when there is missing or flagged data points. 

Traditional calibration software such as AIPS that are based on a scalar data model has the ability to interpolate scalars such as the amplitude or phase of a single polarization. However, this is not possible when we are dealing with $2\times 2$ matrices as what is presented in this paper. Due to the unitary ambiguity, any linear operation in Euclidean space with such matrices would not give us a feasible interpolant. Therefore, we explore the quotient manifold structure \citep{AMS} of the calibration solutions. We use the theory of interpolation over manifolds to get a feasible solution to our problem. We present an algorithm to find the mean of a given set of calibration solutions. This algorithm could also be extended to interpolation with any positive weighting scheme. The interpolant we obtain using this algorithm still has a unitary ambiguity but it is closer to the true interpolant (without ambiguities) than what we obtain by standard (Euclidean) interpolation. Similar work on averaging or interpolation over manifolds and Lie groups appear in diverse areas of research and we refer the reader to \cite{Fiori-Tanaka}, \cite{FioriOptical}, \cite{Kaneko}, and \cite{Amsallem} for further information.

The rest of the paper is organized as follows: In section \ref{calib}, we give an overview of radio interferometric calibration and the ensuing ambiguities. In section \ref{interpolation} we present the interpolation scheme. We provide simulation results to prove the superiority of the proposed scheme in section \ref{simulations} and finally, draw our conclusions in section \ref{conclusions}.

Notation: Matrices and vectors are denoted by bold upper and lower case letters as ${\bf J}$ and ${\bf v}$, respectively. The transpose and the  Hermitian transpose are given by $(.)^T$ and $(.)^H$, respectively. The matrix  Frobenius norm is given by $\|.\|$. The set of complex numbers is denoted by ${\mathbb C}$. The identity matrix is given by $\bf I$. The angle of a complex number is given by $\angle$.

\section[]{Radio interferometric calibration}\label{calib}
Consider a radio interferometer with $N$ stations. Let the sky signal consist of $M$ discrete sources.
The observed data on the baseline $pq$ formed by stations $p$ and $q$ is given by \citep{HBS}
\beq \label{vis}
{\bf V}_{pq} = \sum_{m=1}^M {\bf J}_{pm} {\bf C}_{pqm} {\bf J}_{qm}^H + {\bf N}_{pq}
\eeq
where ${\bf V}_{pq}$ and ${\bf N}_{pq}$  ($\in {\mathbb C}^{2 \times 2}$) are the visibility matrix,  and the noise matrix, respectively.   The instrumental and propagation parameters of stations $p$ and $q$ along the direction of the $m$-th source  are represented by ${\bf J}_{pm}$ and ${\bf J}_{qm}$ ($\in {\mathbb C}^{2 \times 2}$), respectively. The source coherency  matrix of the $m$-th source is given by ${\bf C}_{pqm}$ ($\in {\mathbb C}^{2 \times 2}$). Throughout this paper, we assume all the sources are unpolarized and therefore, ${\bf C}_{pqm}$ to be diagonal. Calibration is the estimation of ${\bf J}_{pm}$ for all $p \in[1,N]$ and $m\in[1,M]$. As noted by \cite{H4}, for any unitary matrix ${\bf U}_m$ ($\in {\mathbb C}^{2 \times 2}$), ${\bf U}_m^H {\bf U}_m = {\bf U}_m {\bf U}_m^H= {\bf I}$, a valid calibration solution would also be ${\bf J}_{pm} {\bf U}_m$. Note that this unitary ambiguity ${\bf U}_m$ can have different values for different $m$, or different directions.

Even with a solution that has an ambiguity, the data could still be corrected for the estimated errors because the ambiguity will cancel out during correction. However, the ambiguity prevents us from using the calibration solutions to model the instrument (such as the beam shape) or propagation phenomena (such as the ones that happen in the ionosphere). The simplest example of further use of solutions is finding their mean. Let us consider having two solutions for station $p$ along the direction $m$, say at adjacent time or frequency intervals: ${\bf J}_{pm1} {\bf U}_{m1}$ and ${\bf J}_{pm2} {\bf U}_{m2}$. Also assume that the true values of the Jones matrices are almost identical, i.e., ${\bf J}_{pm1}\approx {\bf J}_{pm2}={\bf J}_{pm}$. The true mean is $({\bf J}_{pm1}+{\bf J}_{pm2})/2={\bf J}_{pm}$ while we obtain the interpolant $({\bf J}_{pm1} {\bf U}_{m1} + {\bf J}_{pm2} {\bf U}_{m2})/2 = {\bf J}_{pm} (  {\bf U}_{m1} + {\bf U}_{m2})/2$. Due to the fact that $( {\bf U}_{m1} + {\bf U}_{m2})/2$  is not a unitary matrix ($ {\bf U}_{m1}^H  {\bf U}_{m2} \ne {\bf 0}$), the interpolant is not a valid solution for ${\bf J}_{pm}$ anymore and will not satisfy (\ref{vis}).

We can rewrite (\ref{vis}) as 
\beq \label{visJ}
{\bf V}_{pq} = \sum_{m=1}^M {\bf A}_p {\bf J}_{m} {\bf C}_{pqm} {\bf J}_{m}^H {\bf A}_q^T + {\bf N}_{pq}
\eeq
where ${\bf J}_{m}$ is the augmented matrix of all Jones matrices along the $m$-th direction
\beq
{\bf J}_{m}\buildrel\triangle\over= [{\bf J}_{1m}^T,\ldots,{\bf J}_{Nm}^T]^T,\ \in {\mathbb C}^{2N \times 2}.
\eeq
The canonical selection matrix ${\bf A}_p$ (and ${\bf A}_q$ likewise) is given as
\beq \label{Ap}
{\bf A}_p \buildrel\triangle\over=[{\bf 0},{\bf 0},\ldots,{\bf I},\ldots,{\bf 0}],\ \in {\mathbb C}^{2 \times 2N}.
\eeq
In (\ref{Ap}), the $p$-th block of columns is a $2\times 2$ identity matrix while the rest is all $0$. Then, we see that ${\bf J}_m {\bf U}_m$ (where ${\bf U}_m$ is unitary) is a valid solution for ${\bf J}_m$.

In this paper, we try to solve the following problem: Given a set of solutions whose intrinsic values (i.e., solutions without an ambiguity) are almost equal, we find the interpolant that is immune to unitary ambiguities. Let the set of solutions (taken at adjacent time and frequency intervals and even along adjacent directions) be $\mathcal{S}$,
\beq \label{S}
\mathcal{S}=\{{\bf J}_{1}, {\bf J}_{2}, \ldots, {\bf J}_{K}\}
\eeq
that has $K$ elements. The elements in $\mathcal{S}$ satisfy
\beq
{\bf J}_{k}=\widetilde{\bf J}_{k} {\bf U}_{k},\ k\in[1,K]
\eeq
where $\widetilde{\bf J}_{k}$ is the intrinsic value of solution ${\bf J}_{k}$ and ${\bf U}_{k}$ is the unitary ambiguity. We make the additional assumption that all intrinsic values are almost the same, i.e.,
\beq \label{intrinsic_eq}
\widetilde{\bf J}_{1}\approx\widetilde{\bf J}_{2}\ldots\approx\widetilde{\bf J}_{K}.
\eeq
This assumption mostly holds for solutions obtained at adjacent time and frequency intervals as well as along adjacent directions, provided that the scale difference due to source models along adjacent directions is taken into account. We make this assumption so that we are not affected by any aliasing errors. Note that the unitary matrices are almost surely not equal
\beq
{\bf U}_{1}\ne{\bf U}_{2}\ldots\ne{\bf U}_{K}.
\eeq

We restate the problem we need to solve: Given the set of solutions $\mathcal{S}$, find the mean $\overline{\bf J}$ of the solutions such that it is the most accurate approximation of the intrinsic mean
\beq \label{intrinsic}
\overline{\widetilde {\bf J}} = \frac{1}{K}\sum_{k=1}^K \widetilde{\bf J}_{k}
\eeq
within a unitary ambiguity. By replacing the summation with weighted summation in (\ref{intrinsic}), this could also be extended to interpolation with any positive weighting scheme.
\section{Interpolation}\label{interpolation}
In this section, we first present the well established scalar averaging technique used in radio interferometry. As mentioned before, this does not extend to the case with Jones matrices as calibration solutions. In order to proceed further, we give a rather pedagogical overview of the manifold geometry of calibration solutions. Next, we present an averaging algorithm (Algorithm I) using the quotient manifold in section \ref{proposed1}. We also give an alternative algorithm (mainly for comparison purposes) that averages using the tangent space of the quotient manifold  (Algorithm II) in section \ref{proposed2}. 
\subsection{Interpolation: The scalar case}
The calibration solutions of $N$ stations for an interferometer with a single polarization can be given as 
\beq
{\bf g}=\left[ \begin{array}{c}
g_1\\
g_2\\
\vdots\\
g_N 
\end{array} \right] e^{j\psi} =
\left[ \begin{array}{c}
|g_1|e^{j(\angle{g_1+\psi})}\\
|g_2|e^{j(\angle{g_2+\psi})}\\
\vdots\\
|g_N|e^{j(\angle{g_N+\psi})}\\
\end{array} \right] 
\eeq
where $\psi$ is the phase ambiguity common to all stations. Consider the averaging of $K$ such solutions given by the set $\mathcal{G}$ as
\beq
\mathcal{G}=\{ {\bf g}_1 e^{j\psi_1},{\bf g}_2 e^{j\psi_2},\ldots,{\bf g}_K e^{j\psi_K} \}.
\eeq
Consider the calculation of the average phase for station $n$ using the solutions in the set $\mathcal{G}$. Normally, we keep one station (out of $N$) as the reference (say the first station). With one station kept as the reference, the average phase of station $n$ becomes
\beqn \label{scalarav}
\overline{\angle{g_n}}&&=\frac{1}{K} \sum_{k=1}^K (\angle{g_{nk}}+\psi_k-\angle{g_{1k}}-\psi_k) \\\nonumber
&&= \frac{1}{K} \sum_{k=1}^K \angle{g_{nk}} - \frac{1}{K} \sum_{k=1}^K \angle{g_{1k}}
\eeqn
where $\frac{1}{K} \sum_{k=1}^K \angle{g_{nk}}$ is the intrinsic average. Moreover, the term $\frac{1}{K} \sum_{k=1}^K \angle{g_{1k}}$ is common to the averaged phases of all stations. Therefore, phase averaging can be solved for the single polarized case upto a common phase ambiguity. As the ambiguity $e^{j\psi}$ does not affect the amplitudes of the solutions, amplitudes can also be averaged without hindrance. Furthermore, by using positive weights in the summation of (\ref{scalarav}), the same method can be applied to any interpolation scheme. This form of averaging and interpolation is widely used in current interferometric data processing. However, this method does not extend to calibration solutions with dual polarized interferometers where we have Jones matrices as our solutions.

\subsection{The quotient manifold structure of calibration solutions}\label{sec:quotient_manifold}
We provide a brief overview of manifolds and Lie groups before we proceed. A more general overview of this subject can be found in \cite{LTu} and \cite{AMS}.  A manifold can be loosely described as a set of entities, together with a set of mappings (charts) that can locally describe the manifold in Euclidean space. A ``quotient'' manifold is a submanifold of a larger manifold and the entities in the quotient manifold represent more than one entity in the embedding manifold.

This notion of a quotient manifold naturally represents the calibration solutions with unitary ambiguities. Given the set  $\mathcal{S}$ in (\ref{S}), we consider two solutions (say ${\bf J}_{1}$ and ${\bf J}_{2}$)  to be ``similar'' if they are related by a unitary matrix, i.e.,
\beq \label{sim}
{\bf J}_{1} \sim {\bf J}_{2} \Leftrightarrow {\bf J}_{1} = {\bf J}_{2} {\bf U}_{12}
\eeq
where ${\bf U}_{12}$ is unitary. The equivalence relation $\sim$ satisfies reflexive, symmetric, and  transitive conditions as described in \cite{AMS}.  Therefore, assuming the intrinsic value of all elements in $\mathcal{S}$ are the same (\ref{intrinsic_eq}), we can select only one element from $\mathcal{S}$ to represent the whole set, under the equivalence relation $\sim$, given by (\ref{sim}).
\begin{figure}
\begin{minipage}{0.98\linewidth}
\begin{center}
\input{quotient_manifold.pstex_t}\\
\end{center}
\end{minipage}
\caption{The quotient manifold geometry  \citep{AMS} of the calibration solutions. The dashed (blue) line (on $\mathcal{M}$) represents the equivalence class of all solutions that are related to ${\bf J}$ by a unitary ambiguity. This equivalence class is represented by a single point on the quotient manifold $\overline{\mathcal{M}}=\mathcal{M}/\sim$. The vertical space $\mathcal{V}_{\bf J}$ is the vector space tangential to the equivalence class and the horizontal space  $\mathcal{H}_{\bf J}$ is the orthogonal complement.\label{quotient_geom}}
\end{figure}
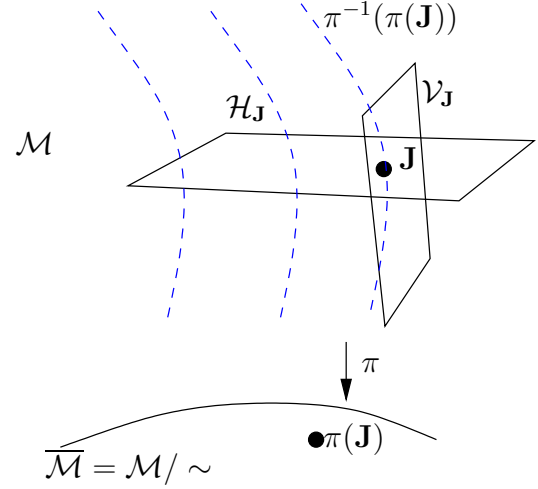

Consider $\mathcal{M}$ to be the manifold of all $2N\times 2$ complex matrices ($\mathbb{C}^{2N\times 2}$), then, we can represent all elements in $\mathcal{S}$ on the quotient manifold $\overline{\mathcal{M}}=\mathcal{M}/\sim$, where the equivalence relation is given by (\ref{sim}). The mapping $\pi$ (canonical projection) is defined such that any matrix ${\bf J}{\bf U}$ on $\mathcal{M}$ (${\bf U}$ unitary) is mapped onto a single point, $\pi({\bf J})$ on $\mathcal{M}/\sim$.

With the mapping $\pi$,  we have the equivalence class
\beq
\pi^{-1}(\pi({\bf J}))=\{ {\bf J}{\bf U}: {\bf U}{\bf U}^H={\bf U}^H{\bf U}={\bf I}, {\bf U}\in \mathbb{C}^{2\times 2}\}
\eeq
of solutions that is represented by a single point on $\mathcal{M}/\sim$, as illustrated in Fig. \ref{quotient_geom}.

The vertical space and horizontal space are vector spaces that are related to the manifold as follows. We take the vertical space to be the tangent space to the equivalence class $\pi^{-1}(\pi({\bf J}))$ at ${\bf J}$, i.e.,
\beq \label{Vspace}
{\mathcal V}_{\bf J}=\{ {\bf J}{\bf A} : {\bf A}^H=-{\bf A}, {\bf A}\in {\mathbb C}^{2\times 2} \}
\eeq
and we choose the orthogonal complement of the vertical space as the horizontal space ${\mathcal H}_{\bf J}$, 
\beq \label{Hspace}
{\mathcal H}_{\bf J}=\{ {\bf J_\bot}{\bf B} : {\bf B}\in {\mathbb C}^{2N-2\times 2} \}.
\eeq
The orthogonal complement ${\bf J}_\bot$ ($\in {\mathbb C}^{2N\times 2N-2}$) is a matrix whose columns are orthogonal to those of the matrix ${\bf J}$, i.e., ${\bf J}^H{\bf J}_\bot={\bf 0}$.

With this formal representation of the manifold geometry of the solution space at hand, we present two algorithms for interpolation in sections \ref{proposed1} and \ref{proposed2}. Similar averaging techniques on other manifolds are already being investigated. For instance, \cite{Kaneko} present an algorithm for averaging on a compact Stiefel manifold. A similar algorithm for averaging on the Grassmann manifold is given by  \cite{Amsallem}. While a manifold has a more geometric structure, a Lie group has a more algebraic structure. A Lie group can be described as a set of entities, with an identity element and operations for multiplication and inverse \citep{LTu}. There is a close relation between smooth manifolds and Lie groups and in \cite{Fiori-Tanaka}, several averaging techniques for square matrices using the Lie group structure is presented. In \cite{FioriOptical}, the Lie group structure as well as the manifold geometry is exploited for averaging.

The algorithm presented in section \ref{proposed1} (Algorithm I), performs the averaging directly on the quotient manifold while the algorithm presented in section \ref{proposed2} (Algorithm II), performs the averaging in the tangent space to the manifold. Algorithm II is presented mainly for comparison with Algorithm I, as most existing work on averaging, such as \citep{Kaneko,Amsallem}, is done in the tangent space. Both these algorithms can be extended to interpolation by using any positive weighting scheme.
\subsection{Averaging on the quotient manifold (Algorithm I)}\label{proposed1}
We present the algorithm to find the mean of the set $\mathcal{S}$ in (\ref{S}). Let us call the estimated mean as $\overline{\bf J}$. The basic idea is to find the element on the quotient manifold that represents the set  $\mathcal{S}$ as accurately as possible. This would also be the mean of the set $\mathcal{S}$.

\begin{enumerate}
\item Start with the initial estimate as one value from $\mathcal{S}$, say $\overline{\bf J} \leftarrow {\bf J}_{1}$. The error threshold is given by $\epsilon$.
\item For each element in $\mathcal{S}$, find unitary ${\bf P}_k$ ($\in {\mathbb C}^{2 \times 2}$) such that
\beq \label{Pk}
 {\bf P}_k=\underset{{\bf P}_k,\ {\bf P}_k^H{\bf P}_k={\bf P}_k{\bf P}_k^H={\bf I}}{\rm arg\ min}\| \overline{\bf J} - {\bf J}_{k}{\bf P}_k \|^2.
\eeq
This is basically an ``alignment'' operation and the details of this step are given in section \ref{minproj}.
\item Form
\beq \label{sum1}
 \overline{\bf G}=\frac{1}{K}\sum_{k=1}^K {\bf J}_{k} {\bf P}_k.
\eeq
\item
Find the unitary projector ${\bf P}$ to minimize $\|\overline{\bf J}-\overline{\bf G} {\bf P}\|^2$ as given in section \ref{minproj}.
\item If $\|\overline{\bf J} - \overline{\bf G} {\bf P}\| < \epsilon$ then stop. Else update $\overline{\bf J} \leftarrow \overline{\bf G}$ and go to step (ii).
\item Return $\overline{\bf J}$ as the mean.
\end{enumerate}

Note that this algorithm can be modified for interpolation by replacing the averages in (\ref{sum1}) with weighted averaging using positive weights. The proof of convergence of this algorithm is part of future research. For the moment, we rely on numerical simulations to test its convergence in section \ref{simulations}.
\subsection{Averaging in the tangent space (Algorithm II)}\label{proposed2}
The algorithm presented in this section projects each element in the set $\mathcal{S}$ to the horizontal space of the quotient manifold before averaging is performed.  As before, let us call the estimated mean as $\overline{\bf J}$. 
\begin{enumerate}
\item Start with the initial estimate as one value from $\mathcal{S}$, say $\overline{\bf J} \leftarrow {\bf J}_{1}$. The error threshold is given by $\epsilon$. For better convergence, a positive scalar $\rho \in (0,1]$ is used.
\item Form the orthogonal projector matrix ${\bf V}$ ($\in \mathbb{C}^{2N \times 2N}$)
\beq
{\bf V}={\bf I}-\overline{\bf J}(\overline{\bf J}^H\overline{\bf J})^{-1}\overline{\bf J}^H.
\eeq
\item Project each element in $\mathcal{S}$, onto the horizontal space $\mathcal{H}_{\overline{\bf J}}$ as
\beq
{\bf W}_k={\bf V} {\bf J}_k.
\eeq
\item Form the average in $\mathcal{H}_{\overline{\bf J}}$ as
\beq \label{sum2}
\overline{\bf W}=\frac{1}{K}\sum_{k=1}^K {\bf W}_k.
\eeq
\item Form current estimate for the average as $\overline{\bf G}= \overline{\bf J}+\rho \overline{\bf W}$. Find the unitary projector ${\bf P}$ to minimize $\|\overline{\bf J}-\overline{\bf G} {\bf P}\|^2$ as given in section \ref{minproj}.
\item If $\|\overline{\bf J} - \overline{\bf G} {\bf P}\| < \epsilon$ then stop. Else update $\overline{\bf J} \leftarrow \overline{\bf G}$ and go to step (ii).
\item Return $\overline{\bf J}$ as the mean.
\end{enumerate}

\subsection{Finding unitary projector to minimize $\| \overline{\bf J} - {\bf J}_{k}{\bf P}_k \|^2$}\label{minproj}
%
What we have to solve is in fact the matrix Procrustes problem \citep{PROCRUS} and we use the algorithm given in \cite{higham}.
\begin{enumerate}
\item Find the product ${\bf X}={\bf J}_{k}^H\overline{\bf J}$.
\item Find  the singular value decomposition of ${\bf X}$ as
\beq
 {\bf U}{\bf S}{\bf V}^H={\bf X}.
\eeq
\item Return ${\bf P}_k={\bf U}{\bf V}^H$.
\end{enumerate}
The proof can be found in \cite{higham}.
\subsection{Discussion}
The main assumption used in Algorithm I is that all elements in $\mathcal{S}$ belong to the equivalence class (dashed line in Fig. \ref{quotient_geom}) or are very close to it. So on the quotient manifold, they lie on a small area that can be considered locally Euclidean, thus enabling averaging. For Algorithm II, we perform the averaging in the tangent space, which is a vector (Euclidean) space and therefore averaging works. Computationally, Algorithm II is much more expensive because of the calculation of the projection matrix at each iteration. Also, it is numerically less stable and hence the need of $\rho$ for better convergence. We use Algorithm II mainly for comparing the performance of Algorithm I. The fundamental question posed here is whether (for our specific problem) it is better to average on the quotient manifold or in the tangent space. Most existing work on averaging or interpolation use operations in the tangent space therefore comparison of Algorithms I and II in terms of their performance is important. In the next section, we give simulation results to compare their performance.

\section[]{Simulation results}\label{simulations}
We consider an interferometric observation with $N=40$ stations and interpolation of $K=10$ solutions. Therefore the set $\mathcal{S}$ in (\ref{S}) has cardinality of $10$, with each matrix ${\bf J}_{k}$ of size $80\times 2$. The intrinsic values of the first matrix, $\widetilde{\bf J}_{1}$, are generated as $\mathcal{U}(0,1)+j\mathcal{U}(0,1)$ (drawn from a uniform distribution in $[0,1]$). The intrinsic values of the  remaining matrices $\widetilde{\bf J}_{2},\ldots,\widetilde{\bf J}_{K}$ are obtained by perturbing the elements of ${\bf J}_{1}$ by adding $\sigma(\mathcal{U}(0,1)+j\mathcal{U}(0,1))$ to them. The value for $\sigma$ is varied for different simulations as explained later. However, we keep the value of $\sigma$ in $[0,0.5]$  to ensure the intrinsic variation is small enough that averaging (or interpolation) is not dominated by aliasing. Once the matrices are generated in this fashion, we multiply each intrinsic  matrix $\widetilde{\bf J}_{k}$ by a random unitary matrix ${\bf U}_{k}$ to get ${\bf J}_{k}=\widetilde{\bf J}_{k} {\bf U}_{k}$. To each realization of ${\bf J}_{k}$, a noise matrix ${\bf N}$ is added. The elements of ${\bf N}$ are generated to be complex circular Gaussian random variables. The values of ${\bf N}$ are scaled to get a given signal to noise ratio (SNR) where the SNR is defined by
\beq \label{SNR}
{\rm{SNR}}\buildrel \triangle \over =\frac{\|{\bf J}_k\|^2}{\|{\bf N}\|^2}
\eeq
before adding them to ${\bf J}_{k}$.

We also calculate the normal (or Euclidean) average for comparison, as
\beq \label{avg}
\widehat{\bf J}=\frac{1}{K}\sum_{k=1}^K {\bf J}_{k}.
\eeq
Moreover, we also calculate the intrinsic sample variance, using the intrinsic mean of (\ref{intrinsic}) as
\beq \label{invar}
\rm{var}(\overline{\widetilde {\bf J}})=\frac{1}{K \|\overline{\widetilde {\bf J}}\|^2} \sum_{k=1}^K \|\overline{\widetilde {\bf J}} -\widetilde{\bf J}_k\|^2.
\eeq

The criterion that we use for measuring the performance of the various averaging algorithms is the normalized error (NE), defined as
\beq \label{NE}
\rm{NE}\buildrel\triangle\over=\frac{\|\overline{\widetilde {\bf J}} -\overline{\bf J} {\bf U}\|^2}{\| \overline{\widetilde {\bf J}}\|^2}
\eeq
where $\overline{\widetilde {\bf J}}$ is the intrinsic mean, $\overline{\bf J}$ is the estimated mean, using (i) Algorithm I (ii) Algorithm II or, (iii) Euclidean average of (\ref{avg}) and finally, ${\bf U}$ is a unitary projector determined as in section \ref{minproj} to minimize $\|\overline{\widetilde {\bf J}} -\overline{\bf J} {\bf U}\|^2$.

\subsection{Simulation I}
\begin{figure}
\begin{minipage}{0.98\linewidth}
\centering
 \centerline{\epsfig{figure=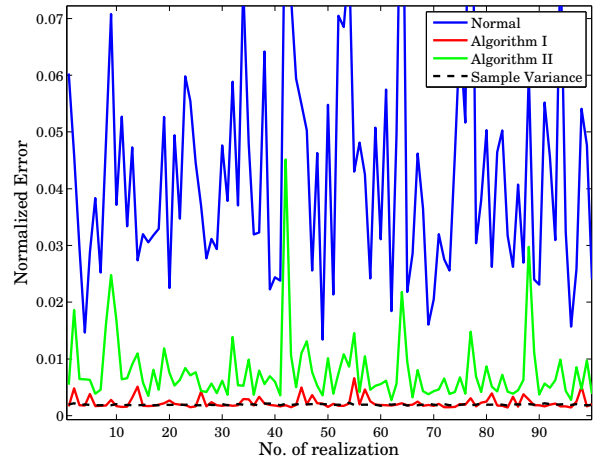,width=9.0cm}}
\caption{Normalized error for $100$ realizations of $\mathcal{S}$ with $\sigma=0.1$ and $\rm{SNR}=100$. The intrinsic sample variance is given by the dashed line. The proposed Algorithm I performs much better than normal (Euclidean) averaging and almost at the level of intrinsic variance. Algorithm II performs better than Euclidean averaging but is worse than Algorithm I.\label{res_norm}}
\end{minipage}
\end{figure}

We generate the set $\mathcal{S}$ as described above $100$ times, keeping $\sigma=0.1$ and $\rm{SNR}=100$. For each realization, we estimate the average by Algorithm I, using $10$ iterations and Algorithm II, using $60$ iterations (the reason for this will be explained later) with $\rho=0.3$. We have shown the normalized estimation error (NE) in Fig. \ref{res_norm} for different approaches. We have also plotted the intrinsic variance, calculated using (\ref{invar}) in this figure. As seen in this figure, the proposed Algorithm I has an  error almost comparable with the intrinsic variance. The proposed Algorithm II has a slightly worse performance but it is still better than Euclidean averaging.

\begin{figure}
\begin{minipage}{0.98\linewidth}
\centering
 \centerline{\epsfig{figure=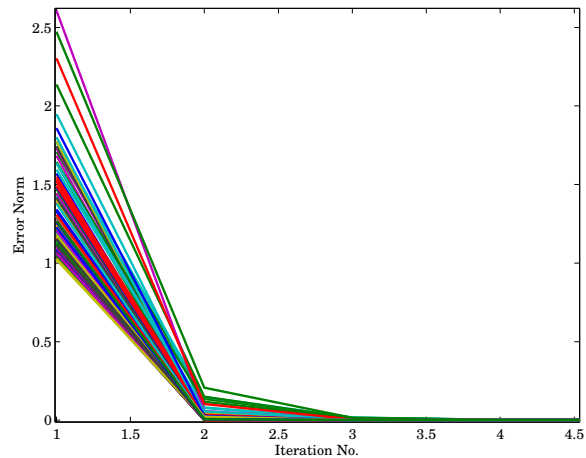,width=9.0cm}}
\caption{The convergence of Algorithm I for $100$ different realizations. After about $4$ iterations, no further improvement occurs and therefore, we can fix the maximum number of iterations at $4$.\label{err_norm_alg1}}
\end{minipage}
\end{figure}

\begin{figure}
\begin{minipage}{0.98\linewidth}
\centering
 \centerline{\epsfig{figure=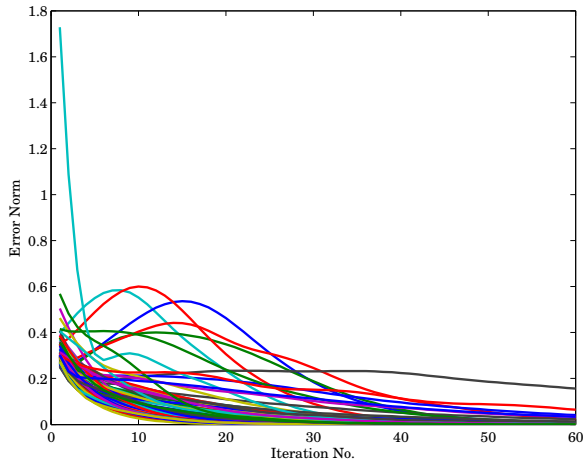,width=9.0cm}}
\caption{The convergence of Algorithm II for $100$ different realizations. Even after $60$ iterations there are some cases that do not converge.\label{err_norm_alg2}}
\end{minipage}
\end{figure}

In Fig. \ref{err_norm_alg1}  and  Fig. \ref{err_norm_alg2}, we have shown the convergence performance of both Algorithm I and Algorithm II. We measure the convergence by the norm of the difference between the  current estimate and the updated estimate at each iteration. It is clear that Algorithm I has much better convergence (only about $4$ iterations) than Algorithm II, which does not converge even after $60$ iterations.

\subsection{Simulation II}
\begin{figure}
\begin{minipage}{0.98\linewidth}
\centering
 \centerline{\epsfig{figure=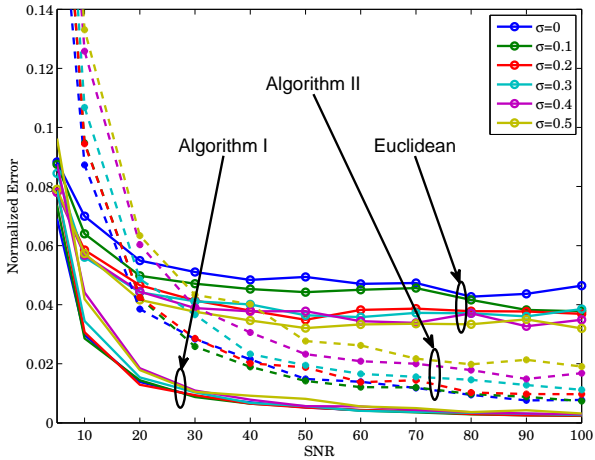,width=9.0cm}}
\caption{The mean normalized error over $100$ realizations for various values of $\sigma$ and SNR. Algorithm I performs significantly better than normal (Euclidean) averaging while Algorithm II performs slightly better but worse than Algorithm I. Algorithm I is almost insensitive to the values of $\sigma$, especially at high values of SNR.\label{err_snr_noweight}}
\end{minipage}
\end{figure}

We vary the values of $\sigma$ and SNR in this simulation. For each value of $\sigma$ and SNR, we generate $100$ realizations of the set $\mathcal{S}$ as before and compute the average. In Fig. \ref{err_snr_noweight}, we have shown the mean normalized error over all realizations for both proposed algorithms as well as for Euclidean averaging. For Algorithm I, we used $4$ iterations and for Algorithm II, we used $60$ iterations with $\rho=0.3$.
From Fig. \ref{err_snr_noweight} it is clear that Algorithm I gives the best results. Moreover, the dependence on $\sigma$ is not significant. The performance of all three methods degrade at low values of SNR. Algorithm II performs worse than Algorithm I and comparing the computational cost, Algorithm I clearly gives better results.

\subsection{Simulation III}
In this simulation, we perform weighted averaging, where the weights are generated from a uniform distribution in $[0,1]$. Once again, for each value of $\sigma$ and SNR, we generate $100$ realizations of $\mathcal{S}$ and for each realization, we generate a new set of weights. We emphasize that Algorithm II did not give convergent results even at $60$ iterations and for any value of $\rho$. Therefore, we omitted the results of Algorithm II because it clearly fails to perform weighted averaging.
\begin{figure}
\begin{minipage}{0.98\linewidth}
\centering
 \centerline{\epsfig{figure=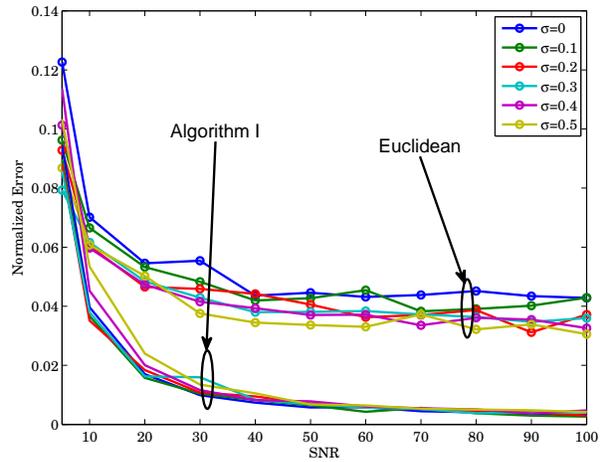,width=9.0cm}}
\caption{The mean normalized  error over $100$ realizations for various values of $\sigma$ and SNR under weighted averaging. Algorithm I performs better than Euclidean averaging while Algorithm II fails to perform well.\label{err_snr_withweight}}
\end{minipage}
\end{figure}

\section[]{Conclusions}\label{conclusions}
We have presented a method for averaging Jones matrices obtained in radio interferometric calibration by exploiting the quotient manifold structure of such solutions. This method could also be used for weighted averaging, or interpolation. Unlike Euclidean averaging, which gives inaccurate results due to the unknown unitary ambiguity in the solutions, the proposed methods give significantly better results. For comparison, we have also proposed an alternative algorithm that operates in the tangent space to the manifold. Simulation results suggest that averaging directly on the quotient manifold is better than averaging by using the tangent space, for our specific problem. One possible reason for this could be due to numerical instability. Existing work that exploit the tangent space for averaging have matrices with orthonormal columns (e.g. the Stiefel manifold). However, in our problem we do not have such a constraint and this could result in numerical instability. Future work would focus on improving the numerical stability and  interpolating along geodesics of the quotient manifold.

\bibliographystyle{mn2e}
\section*{Acknowledgments}
We thank the reviewers: Simone Fiori and Yves Wiaux for a careful review and valuable comments that helped to enhance this paper.

\label{lastpage}

\end{document}

%% file: quotient_manifold.pstex_t
\begin{picture}(0,0)%
\includegraphics{quotient_manifold.pstex}%
\end{picture}%
\setlength{\unitlength}{4144sp}%
\begingroup\makeatletter\ifx\SetFigFont\undefined%
\gdef\SetFigFont#1#2#3#4#5{%
  \reset@font\fontsize{#1}{#2pt}%
  \fontfamily{#3}\fontseries{#4}\fontshape{#5}%
  \selectfont}%
\fi\endgroup%
\begin{picture}(3132,2940)(1696,-4090)
\put(1711,-2086){\makebox(0,0)[lb]{\smash{{\SetFigFont{12}{14.4}{\familydefault}{\mddefault}{\updefault}{\color[rgb]{0,0,0}${\mathcal M}$}%
}}}}
\put(1891,-4021){\makebox(0,0)[lb]{\smash{{\SetFigFont{12}{14.4}{\familydefault}{\mddefault}{\updefault}{\color[rgb]{0,0,0}$\overline{\mathcal M}={\mathcal M}/\sim$}%
}}}}
\put(3556,-3841){\makebox(0,0)[lb]{\smash{{\SetFigFont{12}{14.4}{\familydefault}{\mddefault}{\updefault}{\color[rgb]{0,0,0}$\pi({\bf J})$}%
}}}}
\put(3556,-1321){\makebox(0,0)[lb]{\smash{{\SetFigFont{12}{14.4}{\familydefault}{\mddefault}{\updefault}{\color[rgb]{0,0,0}$\pi^{-1}(\pi({\bf J}))$}%
}}}}
\put(4006,-2176){\makebox(0,0)[lb]{\smash{{\SetFigFont{12}{14.4}{\familydefault}{\mddefault}{\updefault}{\color[rgb]{0,0,0}${\bf J}$}%
}}}}
\put(3781,-3391){\makebox(0,0)[lb]{\smash{{\SetFigFont{12}{14.4}{\familydefault}{\mddefault}{\updefault}{\color[rgb]{0,0,0}$\pi$}%
}}}}
\put(4141,-1771){\makebox(0,0)[lb]{\smash{{\SetFigFont{12}{14.4}{\familydefault}{\mddefault}{\updefault}{\color[rgb]{0,0,0}${\mathcal V}_{\bf J}$}%
}}}}
\put(2971,-1861){\makebox(0,0)[lb]{\smash{{\SetFigFont{12}{14.4}{\familydefault}{\mddefault}{\updefault}{\color[rgb]{0,0,0}${\mathcal H}_{\bf J}$}%
}}}}
\end{picture}%